\documentclass{tlp}
\usepackage{hyperref}
\usepackage{amsmath}
\usepackage{graphicx}
\usepackage{multirow}
\usepackage{asplisting}
\usepackage{times}
\usepackage{url}
\usepackage{todonotes}
\usepackage{listings}
\usepackage{calc}
\usepackage{tikz}
\usetikzlibrary{decorations.pathreplacing}
\usetikzlibrary{shapes.misc}
\usepackage{subfig}
\tikzset{cross/.style={cross out, draw=black, minimum size=3*(#1-\pgflinewidth), inner sep=0pt, outer sep=0pt,thick},
cross/.default={1pt}}

\newlength\listingnumberwidth
\begin{document}

\lefttitle{Tarzariol et al.}

\jnlPage{1}{8}
\jnlDoiYr{2021}
\doival{10.1017/xxxxx}

\title[A CASP-based Solution for Traffic Signal Optimisation]{A CASP-based Solution for Traffic Signal Optimisation}

\begin{authgrp}
\author{\gn{Alice} \sn{Tarzariol}}
\affiliation{AICS, University of Klagenfurt, Austria}\author{\gn{Marco} \sn{Maratea}}
\affiliation{DeMaCS, University of Calabria, Italy}
\author{\gn{Mauro} \sn{Vallati}}
\affiliation{School of Computing and Engineering, University of Huddersfield, UK}
\end{authgrp}

\history{\sub{xx xx xxxx;} \rev{xx xx xxxx;} \acc{xx xx xxxx}}

\maketitle

\begin{abstract}
In the context of urban traffic control, {\sl traffic signal optimisation} is the problem of determining the optimal green length for each signal in a set of traffic signals. The literature has effectively tackled such a problem, mostly with automated planning techniques leveraging the PDDL+ language and solvers. However, such language has limitations when it comes to specifying optimisation statements and computing optimal plans.
In this paper, we provide an alternative solution to the traffic signal optimisation problem based on Constraint Answer Set Programming (CASP). We devise an encoding in a CASP language, which is then solved by means of {\it clingcon 3}, a system extending the well-known ASP solver {\it clingo}. We performed experiments on real historical data from the town of Huddersfield in the UK, comparing our approach to the PDDL+ model that obtained the best results for the considered benchmark. The results showed the potential of our approach for tackling the traffic signal optimisation problem and improving the solution quality of the PDDL+ plans. 
\end{abstract}

\begin{keywords}
Applications, Constraints Answer Set Programming, Urban Traffic Control
\end{keywords}

\section{Introduction}

Urban traffic control aims at minimising average traffic delay in a given region or alleviating the extreme delays of traffic exiting due to major city events and passing through the target area. In this context, \textit{traffic signal optimisation} is the problem of determining the optimal green length for each signal in a set of traffic signals, which may be dispersed around a region consisting of several spatially-close traffic junctions. The problem is usually structured by grouping sets of green signals into stages: each signal in a stage shares the same green time, is situated in the same junction, and collectively lets traffic flow through the junction in a safe manner. This structuring leads to a more convenient representation to solve the problem of determining the optimal green length for each stage.

Typical practical approaches in this context consider fixed-time traffic light phases; thus, with no information about the actual traffic.
Model Predictive Controls (see, e.g., \citep{papageorgiou2013concise}) are hardly used in routine operations, since they are computationally expensive, complicated when it comes to identify the right numerical parameters, and usually slow to converge. 
Traffic-reactive mechanisms are usually deployed in small sets of neighbouring junctions (between three and nine), and can take decisions in real-time on how to adapt stage duration based on sensor data \citep{SCOOT}. A major issue comes from the fact that reactive methods can not leverage on knowledge of incoming traffic that has not yet hit the controlled region, or on wider information about traffic in the area. 
To support this sort of reasoning, the traffic signal optimisation problem has been more recently tackled in the literature with automated planning techniques \citep{DBLP:journals/aim/Smith20,DBLP:conf/itsc/VallatiC23} leveraging PDDL+ language and solvers \citep{vallati2016efficient,antoniou2019enabling,percassi2023practical}, then extended to cope with a legacy traffic control infrastructure taking into account deployment constraints \citep{DBLP:conf/icaps/KouaitiPSMV24}. The automated planning solutions demonstrated interesting capabilities, and the generated strategies have been deployed in real-world trials in the Hull city region and in the Kirklees council area in the United Kingdom. 

Despite the successful deployments, PDDL+ has limitations when it comes to specifying optimisation statements and to computing optimal plans, which are particularly useful in practice to ensure that benefits materialise in the controlled region as soon as possible. 
At the state of the art, PDDL+ planning engines focus on generating satisfying solutions, with no guarantees on the quality of the solution found. In practice, this can lead to traffic signal optimisation plans that show extremely long time horizons -- potentially reducing the ability of plans to cope with identified issues.
Considering such limitations, the nature of the problem and the additional constraints and domain size of the actual real-world infrastructure of the setting by \cite{DBLP:conf/icaps/KouaitiPSMV24}, in
this paper we present an alternative, novel solution to the traffic signal optimisation problem based on Constraint Answer Set Programming (CASP)~\citep{DBLP:journals/amai/MellarkodGZ08,DBLP:journals/tplp/BalducciniL17,DBLP:conf/kr/LiuJN12,DBLP:journals/tplp/JanhunenKOSWS17}, which integrates ASP and Constraint Programming. 

Our encoding can model the same solution space of the  PDDL+ models from \cite{DBLP:conf/icaps/KouaitiPSMV24}, but limited up to a certain time horizon.
After observing that the (pure) ASP representation struggles to scale with meaningful horizons for this problem, we extend it with CASP statements in order to tackle producing plans of higher horizons.
In particular, we apply the system {\it clingcon 3} \citep{clingcon}, which is an implementation that extends the well-known ASP solver {\it clingo} with theory atoms and propagators for linear constraints. 
For the experiments, we compare and evaluate our approach with respect to \textsc{FiRe}, the PDDL+ model that obtained the best performance among the proposed alternatives in \citep{DBLP:conf/icaps/KouaitiPSMV24} and that, on average, showed better plans than the ones used in historical data. 
The benchmark is drawn from real data 
and we analysed two tasks: first, evaluating whether \textit{clingcon} is capable of finding or improving solutions with certain restrictions; and second, evaluating the result of   \textit{clingcon} with optimisation statements. 
The results showed that our CASP encoding is a promising approach to tackle the traffic signal optimisation problem with limited horizons.

The paper is structured as follows: Section \ref{sec:prel} introduces needed preliminaries about ASP and \textit{clingcon 3} language. Then, Section \ref{sec:desc} describes the traffic signal optimisation problem we address in this paper, while our CASP encoding is presented in Section \ref{sec:enc}. Further, Section \ref{sec:exp} shows the results of an experimental analysis comparing our new approach to the best one in \citep{DBLP:conf/icaps/KouaitiPSMV24}, and proposes possible alternative objectives that exploit the optimisation capabilities of \textit{clingcon}. 
The paper ends in Section \ref{sec:rel} and \ref{sec:conc} by discussing related literature and drawing final remarks, respectively. 

\section{Preliminaries}
\label{sec:prel}
\paragraph{Answer Set Programming (ASP)} \citep{DBLP:journals/ngc/GelfondL91,baral2003,DBLP:journals/amai/Niemela99,DBLP:journals/cacm/BrewkaET11} 
is a declarative programming paradigm that applies non-monotonic reasoning and relies on the stable model semantics. 
In the following, we describe a fragment of the ASP syntax, focusing on the constructs appearing in our encoding. An ASP program $P$ is a set of \emph{rules}~$r$ of the form: 
\lstinline{h}\lstinline{ :-  b}$_1$\lstinline{,}$\dots$\lstinline{,b}$_n$\lstinline{.},
 where \lstinline{h} is an atom and each \lstinline{b}$_i$, for $1\leq i \leq n$, is a literal, a comparison, or an aggregate. 
An \emph{atom} is an expression of the form \lstinline{p(t}$_1$\lstinline{,}$\dots$\lstinline{,t}$_m$\lstinline{)}, where \lstinline{p} is a predicate and \lstinline{t}$_i$, for $1\leq i \leq m$, are terms. 
A \emph{literal} \lstinline{c} is either an atom \lstinline{a}$_i$ or its negation \lstinline{not a}$_i$. 
A comparison is equal to \lstinline{t}$_1\circ{}$\lstinline{t}$_2$
where \lstinline{t}$_1$ and \lstinline{t}$_2$ are terms and $\circ\in\{\text{\lstinline{<=}},\text{\lstinline{<}},\text{\lstinline{=}},\text{\lstinline{>}},\text{\lstinline{>=}}\}$.
An aggregate is of the form \lstinline{t}\lstinline|=#count{t|$_1$\lstinline{:c}$_1, \cdots$,\lstinline{ t}$_n$\lstinline{:c}$_n$\lstinline|}| 
or \lstinline{t}\lstinline|=#sum{v|$_1$,\lstinline|t|$_1$\lstinline{:c}$_1, \cdots, $\lstinline{ v}$_n$,\lstinline{t}$_n$\lstinline{:c}$_n$\lstinline|}| 
where \lstinline{t} and each \lstinline{v}$_i$ are integers, \lstinline{t}$_i$ are terms and \lstinline{c}$_i$ are literals, for $1\leq i \leq n$.
The program $P$ can also contain \emph{choice rules}~$r$ of the form: 
\lstinline|{a|$_1$\lstinline{:c}$_1$\lstinline{;}$ \cdots$\lstinline{;a}$_n$\lstinline{:c}$_n$\lstinline|}=s|\lstinline{ :-  b}\lstinline{.},
 where \lstinline{b} is an atom,  \lstinline{s} is a positive integer and, for $1\leq i \leq n$, \lstinline{a}$_i$ is an atom and \lstinline{c}$_i$ is a literal.
A rule~$r$ is called a \emph{fact} when $n=0$, and a \emph{constraint} if \lstinline{h} is not present.
 The left side of the symbol \lstinline{:-} in a rule is called \textit{head}, while the right side is called \textit{body}.
The semantics of an ASP program $P$ is given in terms of the answer sets of its \emph{ground instantiation} $P_{grd}$, computed by replacing each (first-order) rule $r\in\nolinebreak P$ with ground rules obtained by substituting the variables in~$r$ by constants occurring in~$P$ and evaluating arithmetical expressions.
An \textit{answer-set} is a collection of (true) ground atoms such that all rules of $P_{grd}$ are satisfied and allow for deriving each of the ground atoms in the head of some rule whose body is satisfied.
We refer to \cite{CalimeriFGIKKLM20} for more details on the ASP syntax and semantics.

\paragraph{Clingcon.} Among the various CASP dialects, we focus in this paper on the one of \textit{clingcon 3} \citep{clingcon}, which is the CASP solver we used in the experiments. It extends the fifth generation of the ASP system \textit{clingo} by introducing theory atoms in its language and augmenting the solver with propagators for linear constraints, allowing it to deal with integer domains of considerable size.
Among its language extensions, \textit{clingcon 3} defines theory atoms as expressions such as  \lstinline|&dom{l..u}=a|, where \lstinline{l} and \lstinline{u} are integers and \lstinline{a} is a constraint variable, or expressions as \lstinline|&sum{t1;t2;...;tn}=t0|, where each \lstinline{ti} for \lstinline{i} $\in [$\lstinline{0}$..$\lstinline{n}$]$ is a constraint variable or an integer. After grounding, the constraint expressions of the former type are transformed into domain restrictions, requiring that the value associated to \lstinline{a} is an integer included in $[$\lstinline{l}$..$\lstinline{u}$]$, while the latter become linear constraint atoms requiring $\sum_{i\in[1..n]}$\lstinline{ti}$=$ \lstinline{t0}.
Lastly, expressions as \lstinline|&maximize{a}| and \lstinline|&minimize{a}| where \lstinline{a} is a constraint variable become directives that define as optimal any answer set with the maximal or minimal possible assignment for \lstinline{a}, respectively.

\section{Scenario and Problem Description}
\label{sec:desc}

\begin{figure}
    \centering
    \includegraphics[width=0.7\linewidth]{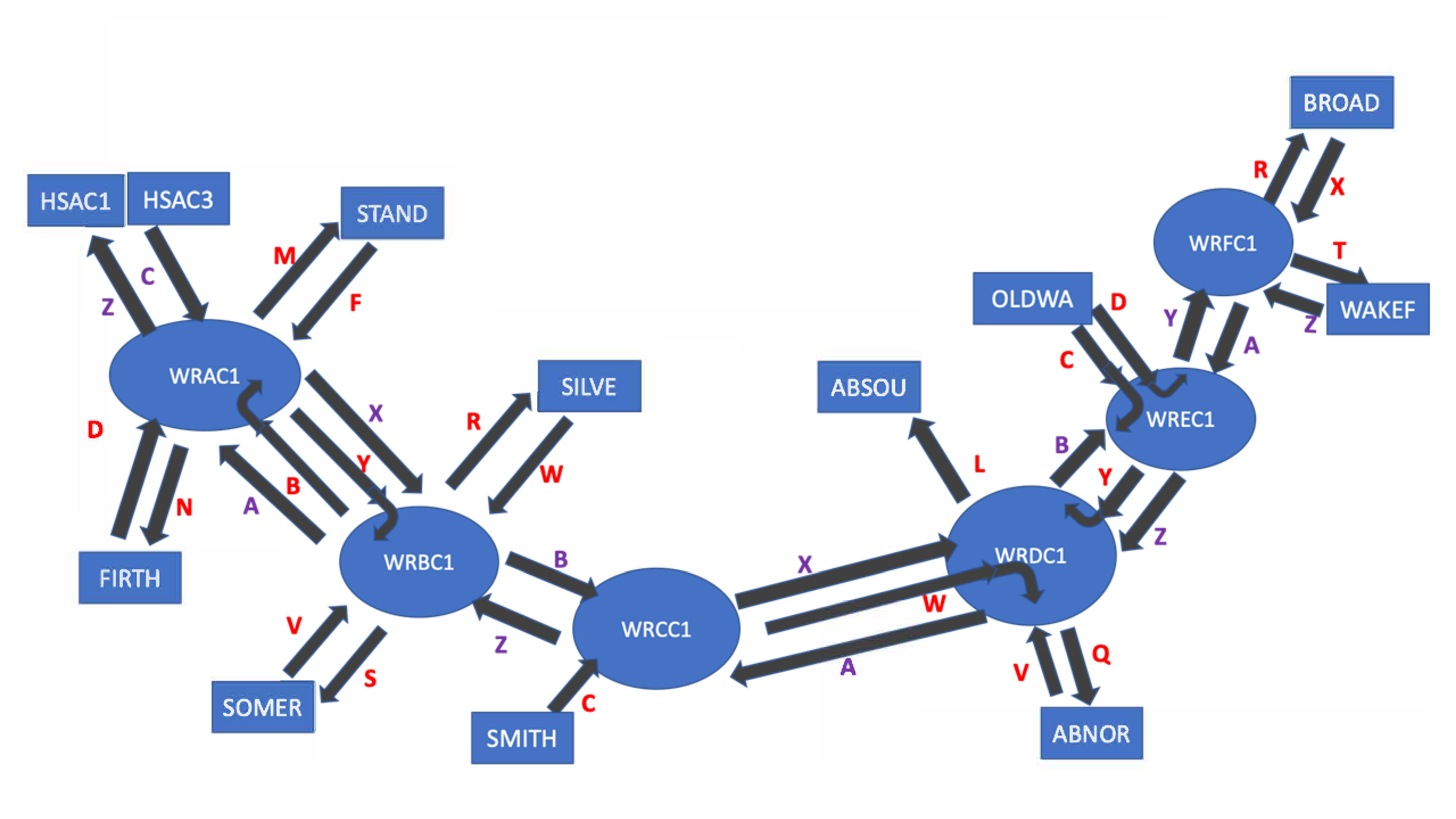}
    \caption{A diagram of the considered corridor in terms of junctions (circles), links, and boundaries (rectangles). For readability, the map is not correctly scaled.}
    \label{fig:map}
\end{figure}

This section describes the elements and constraints considered in the traffic signal optimisation problem, in particular using the same formalism and scenario applied in \citep{DBLP:conf/icaps/KouaitiPSMV24}. It is worth reminding that this formalism is based on a mesoscopic representation of traffic \citep{ferrara2018microscopic}, in which the number of vehicles in road links are considered rather than representing individual vehicles. This is a standard approach to reduce complexity and ensure the solvability of the problems. 

The general goal of traffic signal optimisation is to minimise the average traffic delay for a region of interest. In some cases, it can be more specific and focus on alleviating the extreme delays of traffic exiting major city events and passing through the region, or dealing with accidents or unusual events. 
In this work, we focus on a major corridor in the Kirklees council area within West Yorkshire, United Kingdom, which is approximately 1.3 kilometres long. 
The corridor, shown in Figure \ref{fig:map} in a simplified form, allows traffic from the Huddersfield ring road (that sits on the left of the corridor) to reach the M62 and M1 highways (right-hand side), and vice versa. Further, it also serves people joining or leaving events hosted at the nearby John Smith’s Stadium. 
The corridor includes 6 \textit{junctions} and considers 34 road \textit{links}. 

Each junction contains traffic signals, the status of which determines whether vehicles on certain (incoming) roads are allowed to proceed through to other (outgoing) ones. In the problem representation, the concept of traffic signals in a junction is abstracted, and instead we use the concept of the junction's \textit{stage}, i.e., set of traffic movements that can be active at the same time on the junction, hence controlling the flows of vehicles between connected links.
A \textit{cycle} is a complete sequence of all stages, defining their order. In our scenario, stages are always interleaved by \textit{intergreen} times, i.e., times where no traffic movement occurs in the junction. This occurs when all the junction's traffic signals are on red to allow pedestrians to cross and vehicles still transiting in the junction to leave the area before the next stage begins. 
Constraints on the problem include the legal and practical restrictions on the minimum and maximum duration of each stage, and the minimum and maximum length of the overall traffic signal cycle. The order of stages in a cycle can not be modified. 

We can then formalise the problem's objective as optimising the length of traffic signal stages for each junction in the controlled urban region, to minimise the average traffic delay.
In the focus area, the SCOOT \citep{SCOOT} system is in operation for performing traffic signal optimisation. SCOOT is a traffic reactive control mechanism used widely around the world, and is aimed at handling cycle-to-cycle changes in traffic demand. 
In response to changes in traffic flows, SCOOT would gradually adapt the traffic signal timings of a set of managed neighbouring junctions. The adapting process is gradual, in the range of 4-8 seconds difference per cycle, and naturally discretised: the minimum granularity is 1 second. In performing its task, SCOOT is dependent on its own local data sensors, usually inductive loops embedded in the road surface, and stores sensed data and operational information in a dedicated database. 

Exploiting the architecture proposed by \cite{DBLP:conf/ijcai/BhatnagarGMMPV23}, we can extract information from the SCOOT infrastructure and simulate historical data and generated solutions. 
Such architecture allows the use of external tools to perform traffic signal optimisation, and to inject the generated strategies to be deployed in the region. 
To allow deployment on heritage traffic control infrastructure \citep{DBLP:conf/icaps/KouaitiPSMV24}, the traffic signal optimisation problem needs to be redefined by considering the following additional constraints: (i) the length of the stages can not be modified arbitrarily; instead, for each junction, the \textit{configuration} of cycles (i.e., the specification of the length of every stage in the cycle) can only
be selected from a predefined set, and (ii) the cycles considered for the junctions in the controlled region should have approximately the same duration -- to avoid synchronisation issues and ensure that green waves are preserved. 

The problem formalisation applied in \citep{DBLP:conf/icaps/KouaitiPSMV24}  abstracts the vehicle capacity of all the road links by considering instead the numbers of ``passenger car units" (PCU), which is the standard unit for measuring traffic flows, corresponding to the typical passenger car. 
Moreover, it represents with the \textit{turnrate} the average traffic flows between links in number of PCU's per second, i.e., the number of
vehicles flowing through a particular junction at a certain time of day, when the corresponding traffic signal stage is green. 
Lastly, to minimise the average traffic delay, the problem formalisation uses the 
 concept of \textit{counter} to measure the number of vehicles that navigated through
the link over a considered period of time, and was introduced in \cite{DBLP:conf/mtits/PercassiBGMMV23} to support the heuristic reasoning of the planning engine. Increasing the number of vehicles that navigate the region in a given period of time is a proxy for minimising average delay, as it aims at increasing the throughput of the network, hence reducing time wasted queuing. 
In this work, we follow the same concept.

\section{CASP Modelling}
\label{sec:enc}
This section introduces our representation of the traffic signal optimisation problem by means of ASP and \textit{clingcon 3} language. We start by illustrating the facts contained in each problem instance in Subsection \ref{subsec:instance}, and then we introduce our encoding. 
To improve the readability, we split it into two parts, first describing the rules where exclusively ASP is used, followed by the part where \textit{clingcon} expressions appear. 
The Subsection \ref{subsec:enc1} models the decision points of the solver, i.e., the time where a configuration can be selected for each junction, and the status of the junctions at each time, i.e., which stage or intergreen time is active. 
On the other hand, Subsection \ref{subsec:enc2}  uses \textit{clingcon} expressions to model the occupancy, i.e., the number of PCU present at each moment, and the counter of each link, i.e., the number of PCU entering it. 
Differently from the PDDL+ models, where the heuristics of the planner are used to return a promising sequence of configurations to reach a target value for every counter, with our ASP model we can define an optimisation statement aiming to maximise the values of counters within a given horizon.

\subsection{Problem Instances}
\label{subsec:instance}
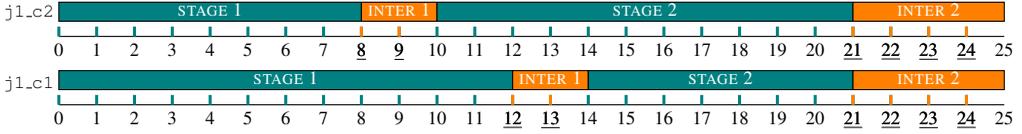
\begin{figure}
\centering
    \begin{tikzpicture}[%
    every node/.style={
        font=\scriptsize,
        text height=1ex,
        text depth=.25ex,
    },
]

\node[anchor=north] at (-0.4,0.47) {\texttt{j1\_c1}};
\node[anchor=north] at (-0.4,1.41) {\texttt{j1\_c2}};
\draw[-] (0,0) -- (12.5,0);
\draw[-] (0,0.9) -- (12.5,0.9);

\foreach \x in {0,1,...,24}{
    \draw[line width=0.4mm, teal] (\x/2
    ,0) -- (\x/2,0.13);
    \draw[line width=0.4mm, teal] (\x/2,0.9) -- (\x/2,1.03);
    \node[anchor=north] at (\x/2,0) {\x};
    \node[anchor=north] at (\x/2,0.9) {\x};
}
  \node[anchor=north] at (25/2,0) {25};
   \node[anchor=north] at (25/2,0.9) {25};

\foreach \x in {12,13,21,22,23,24}{
    \draw[line width=0.4mm, orange] (\x/2,0) -- (\x/2,0.13);
    \node[anchor=north] at (\x/2,0) {\underline{\x}};
}

\foreach \x in {8,9,21,22,23,24}{
    \draw[line width=0.4mm, orange] (\x/2,0.9) -- (\x/2,1.03);
    \node[anchor=north] at (\x/2,0.9) {\underline{\x}};
}

\draw[black,fill=teal] (0,0.2) rectangle (6,0.45)node[midway,text=white]{\textsc{stage 1}};
\draw[black,fill=orange] (6,0.2) rectangle (7,0.45)node[midway,text=white]{\textsc{inter 1}};
\draw[black,fill=teal] (7,0.2) rectangle (10.5,0.45)node[midway,text=white]{\textsc{stage 2}};
\draw[black,fill=orange] (10.5,0.2) rectangle (12.5,0.45)node[midway,text=white]{\textsc{inter 2}};

\draw[black,fill=teal] (0,1.1) rectangle (4,1.35)node[midway,text=white]{\textsc{stage 1}};
\draw[black,fill=orange] (4,1.1) rectangle (5,1.35) node[midway,text=white]{\textsc{inter 1}};
\draw[black,fill=teal] (5,1.1) rectangle (10.5,1.35)node[midway,text=white]{\textsc{stage 2}};
\draw[black,fill=orange] (10.5,1.1) rectangle (12.5,1.35)node[midway,text=white]{\textsc{inter 2}};

\end{tikzpicture}
    \vspace*{-2mm}
\caption{Example of cycles of $25$ seconds with two stages for two configurations, \lstinline{j1_c1} and \lstinline{j1_c2}.}
    \label{fig:cycle}
\end{figure}

\begin{lstlisting}[float=b,label=prg:cycle,numbers=none,caption={ASP facts describing the cycle in Figure \ref{fig:cycle}.}]
available_conf(j1,j1_c1).            available_conf(j1,j1_c2).           
next(stage(j1,1),inter(j1,1)).       next(stage(j1,2),inter(j1,2)). 
next(inter(j1,1),stage(j1,2)).       next(inter(j1,2),stage(j1,2)). 
status(j1,stage(j1,1)).        status(j1,inter(j1,1)). 
status(j1,stage(j1,2)).        status(j1,inter(j1,2)).      end(inter(j1,2)).
phase_limit(stage(j1,1),j1_c1,12).   phase_limit(inter(j1,1),j1_c1,2).
phase_limit(stage(j1,2),j1_c1,7).    phase_limit(inter(j1,2),j1_c1,4).
phase_limit(stage(j1,1),j1_c2,8).    phase_limit(inter(j1,1),j1_c2,2).
phase_limit(stage(j1,2),j1_c2,11).   phase_limit(inter(j1,2),j1_c2,4).
\end{lstlisting}
Problem instances represent the initial setting and status of the corridor to be managed, containing a set of facts defining the following atoms: 
\lstinline{controllable(J)} identifies every junction \lstinline{J} for which the solver can change its stages' green light duration, selecting from the set of available configurations, defined through the atoms \lstinline{available_conf(J,C)}. 
In our encoding, a cycle is represented as a sequence of phases, where a phase is either a traffic light stage or an intergreen time. Specifically, for each configuration \lstinline{C} the set of atoms \lstinline{phase_limit(P,C,D)} defines the duration \lstinline{D} of the phase \lstinline{P} (where \lstinline{P} is either a stage or an intergreen).  
The atoms of the form \lstinline{status(J,P)} specify the 
phases \lstinline{P} occurring in a junction \lstinline{J}; the order of phases in a cycle is defined through the atoms \lstinline{next(P1,P2)}, where the phase \lstinline{P1} is followed by \lstinline{P2}, and \lstinline{end(P)} identifies the final intergreen time. 
Listing \ref{prg:cycle} contains a simple example of ASP atoms characterising the junction cycle in Figure \ref{fig:cycle}.

The atoms of the form \lstinline{link(J1,ID,J2)} represent (directed) road links connecting the junction \lstinline{J1} to \lstinline{J2} and are identified by \lstinline{ID} to avoid ambiguities for the same connections. 
Thus, we can represent the top left links in Figure \ref{fig:map} as, for example,  \lstinline{link(wrac1,z,hsac1)} and \lstinline{link(hsac1,c,wrac1)}.
For the sake of compactness, we use the unary predicate \lstinline{link(L)}, where \lstinline{L} is equal to \lstinline{link(J1,ID,J2)}.
Moreover, the atoms \lstinline{precedes(J,L)} and \lstinline{follows(J,L)} identify the junction \lstinline{J} preceding and following each link \lstinline{L}, respectively. 
Every link \lstinline{L} has a maximum capacity and initial occupancy, contained in the second term of \lstinline{capacity(L,C)} and \lstinline{initial_occ(L,O)}, respectively. Capacity defines the maximum amount of PCU that can be in the link at the same time, while occupancy indicates the actual amount of PCU in the link in the beginning of the simulation. If the atom capacity is not defined for a link, then we assume that there is no limit for it.
Moreover, the atoms \lstinline{turnrate(S,L1,L2,U)} contain the PCU \lstinline{U} that transits from a link \lstinline{L1} to link \lstinline{L2} during the stage \lstinline{S} per unit of time (i.e., second). 
The capacity, occupancy and turn rate numbers have a precision of five decimal digits and are normalised by multiplying each number by $10^5$.
The initial status of the corridor (i.e., time 0) is described by (i) the active phase in each junction, entailed by the second term of the atoms \lstinline{active_p(0,P)} (with just one active stage \lstinline{P} per junction), (ii) the amount of time since \lstinline{P} is active, entailed by the third term of the atoms \lstinline{active_t(0,J,T)} (note that \lstinline{J} is the junction where \lstinline{P} is active), and (iii) the active configuration for each junction, defined through the third term of \lstinline{active_c(0,J,A)}.
%
Lastly, the atoms \lstinline{initial_count(L,D)} contain the links \lstinline{L} for which we want to maximise the flow of vehicles, where \lstinline{D} is the initial value that we assign to the counter (by default, it is equal to zero).

\subsection{ASP Encoding}
\label{subsec:enc1}
\begin{lstlisting}[float=b,label=prg:encoding,caption={Encoding part 1 - Define decision points and set configuration}.]
cycle(J,D) :- available_conf(J,C), D=#sum{T,P: phase_limit(P,C,T)}.

prev_status(S,S1) :- next(S1,S), not endcycle(S1).
prev_status(S,S2) :- prev_status(S,S1), next(S2,S1), not end(S2).

sub(J,T+M):- active_conf(0,J,A), status(J,S), active_t(0,J,T), active_p(0,S),
             M=#sum{L1,S1: prev_status(S,S1), phase_limit(S1,A,L1)}.
step(J,1,D-M) :- controllable(J), cycle(J,D), sub(J,M), D-M<=horizon. 
step(J,C,T+D) :- step(J,C-1,T), cycle(J,D), T+D<=horizon, C>1.

{conf(J,C,T,A): available_conf(J,A) } = 1 :- step(J,C,T).
conf(J,0,0,A) :- active_c(0,J,A).

change(J,C,T,A) :- conf(J,C,T,A), 0<C, not conf(J,C-1,_,A).
:- conf(J,0,0,A), count_c(J,I), C=1..k-I-1, not conf(J,C,T1,A), step(J,C,T1).
:- change(J,C,T,A), I=1..k-1, not conf(J,C+I,T1,A), step(J,C+I,T1).
\end{lstlisting}
Our ASP representation simulates the status of the corridor at every second, from time $0$ up to 
the \lstinline{horizon}. 
The unique choice rule in our encoding selects the active configuration for each junction. 
The decision points can be precomputed since changing configuration during a cycle is not allowed and the length of different configurations must coincide, as specified in Section \ref{sec:desc}.
Listing \ref{prg:encoding} contains the rules to determine the decision points and configuration for every junction. 
The rule in line $1$ computes in \lstinline{cycle(J,D)} the duration \lstinline{D} of a cycle for a junction \lstinline{J}. 
The atoms \lstinline{prev_status(P,P1)} enumerate, for each phase \lstinline{P}, all its previous phases \lstinline{P1} up to the beginning of the cycle. The rule in line $6$ calculates with the atoms \lstinline{sub(J,S)} the seconds \lstinline{S} since the current cycle in \lstinline{J} was active, up to the point where the simulation has begun. The rules in lines $8$ and $9$ use these atoms to derive  \lstinline{step(J,C,T)}, which identifies for each controllable junction \lstinline{J} the time point \lstinline{T} in which the \lstinline{C}-th cycle ends; in other words, \lstinline{T} is the \lstinline{C}-th time point where the configuration of \lstinline{J} can be changed.
Lastly, line $11$ contains the choice rule for \lstinline{conf(J,C,T,A)} that selects the active configuration \lstinline{A} for each atom \lstinline{step(J,C,T)}, while its value at time $0$ is defined in line $12$.
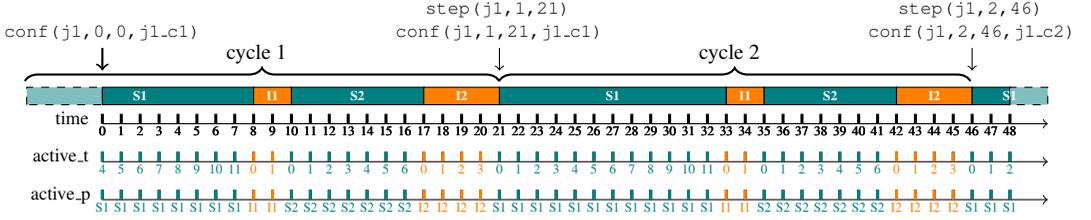
\begin{figure}
\centering
    \begin{tikzpicture}[%
    every node/.style={
        font=\scriptsize,
        text height=1ex,
        text depth=.25ex,
    },
]

\foreach \y in {0,-0.5,-1}{
\draw[->] (0,\y) -- (12.5,\y);
}
\node at (-0.4,0.05) {time};
\node at (-0.55,-0.45) {active\_t};
\node at (-0.55,-0.95) {active\_p};
\foreach \y in {0,-0.5,-1}{
\foreach \x in {0,1,...,48}{
    \draw[line width=0.4mm] (\x/4,\y) -- (\x/4,\y+0.13);
    \node[anchor=north] at (\x/4,0.1) {\tiny \x};
}
 
}

\foreach \x in {0,1,...,7}{
    \node[anchor=north] at (\x/4,-0.4) {\tiny \textcolor{teal}{\number\numexpr\x+4\relax}};
    \node[anchor=north] at (\x/4,-0.9) {\tiny \textcolor{teal}{S1}};
    \draw[line width=0.4mm, teal] (\x/4,-0.5) -- (\x/4,-0.37);
    \draw[line width=0.4mm, teal] (\x/4,-1) -- (\x/4,-0.87);
}
\foreach \x in {8,9}{
    \node[anchor=north] at (\x/4,-0.4) {\tiny \textcolor{orange}{\number\numexpr\x-8\relax}};
    \node[anchor=north] at (\x/4,-0.9) {\tiny \textcolor{orange}{I1}};
           \draw[line width=0.4mm, orange] (\x/4,-0.5) -- (\x/4,-0.37);
    \draw[line width=0.4mm, orange] (\x/4,-1) -- (\x/4,-0.87);
}
\foreach \x in {10,...,16}{
    \node[anchor=north] at (\x/4,-0.4) {\tiny \textcolor{teal}{\number\numexpr\x-10\relax}};
    \node[anchor=north] at (\x/4,-0.9) {\tiny \textcolor{teal}{ S2}};
    \draw[line width=0.4mm, teal] (\x/4,-0.5) -- (\x/4,-0.37);
    \draw[line width=0.4mm, teal] (\x/4,-1) -- (\x/4,-0.87);
}
\foreach \x in {17,18,19,20}{
    \node[anchor=north] at (\x/4,-0.4) {\tiny \textcolor{orange}{\number\numexpr\x-17\relax}};
    \node[anchor=north] at (\x/4,-0.9) {\tiny \textcolor{orange}{I2}};
       \draw[line width=0.4mm, orange] (\x/4,-0.5) -- (\x/4,-0.37);
    \draw[line width=0.4mm, orange] (\x/4,-1) -- (\x/4,-0.87);
}
\foreach \x in {21,...,32}{
    \node[anchor=north] at (\x/4,-0.4) {\tiny \textcolor{teal}{\number\numexpr\x-21\relax}};
    \node[anchor=north] at (\x/4,-0.9) {\tiny \textcolor{teal}{S1}};
       \draw[line width=0.4mm, teal] (\x/4,-0.5) -- (\x/4,-0.37);
    \draw[line width=0.4mm, teal] (\x/4,-1) -- (\x/4,-0.87);}
    
\foreach \x in {33,34}{
    \node[anchor=north] at (\x/4,-0.4) {\tiny \textcolor{orange}{\number\numexpr\x-33\relax}};
    \node[anchor=north] at (\x/4,-0.9) {\tiny \textcolor{orange}{I1}};
       \draw[line width=0.4mm, orange] (\x/4,-0.5) -- (\x/4,-0.37);
    \draw[line width=0.4mm, orange] (\x/4,-1) -- (\x/4,-0.87);
}
\foreach \x in {35,...,41}{
    \node[anchor=north] at (\x/4,-0.4) {\tiny \textcolor{teal}{\number\numexpr\x-35\relax}};
    \node[anchor=north] at (\x/4,-0.9) {\tiny \textcolor{teal}{S2}};
       \draw[line width=0.4mm, teal] (\x/4,-0.5) -- (\x/4,-0.37);
    \draw[line width=0.4mm, teal] (\x/4,-1) -- (\x/4,-0.87);
}

\foreach \x in {42,43,...,45}{
    \node[anchor=north] at (\x/4,-0.4) {\tiny \textcolor{orange}{\number\numexpr\x-42\relax}};
    \node[anchor=north] at (\x/4,-0.9) {\tiny \textcolor{orange}{I2}};
       \draw[line width=0.4mm, orange] (\x/4,-0.5) -- (\x/4,-0.37);
    \draw[line width=0.4mm, orange] (\x/4,-1) -- (\x/4,-0.87);
}
\foreach \x in {46,...,48}{
    \node[anchor=north] at (\x/4,-0.4) {\tiny \textcolor{teal}{\number\numexpr\x-46\relax}};
    \node[anchor=north] at (\x/4,-0.9) {\tiny \textcolor{teal}{S1}};
       \draw[line width=0.4mm, teal] (\x/4,-0.5) -- (\x/4,-0.37);
    \draw[line width=0.4mm, teal] (\x/4,-1) -- (\x/4,-0.87);
}

\draw[black,dashed,fill=teal!50] (-1,0.25) rectangle (0,0.48);
\draw[black,fill=teal] (0,0.25) rectangle (2,0.48);
\draw[fill=none,draw=none] (-1,0.25) rectangle (2,0.48)node[midway,text=white]{\tiny \textbf{S1}};
\draw[black,fill=orange] (2,0.25) rectangle (2.5,0.48)node[midway,text=white]{\tiny \textbf{I1}};
\draw[black,fill=teal] (2.5,0.25) rectangle (4.25,0.48)node[midway,text=white]{\tiny \textbf{S2}};
\draw[black,fill=orange] (4.25,0.25) rectangle (5.25,0.48)node[midway,text=white]{\tiny \textbf{I2}};
\draw[black,fill=teal] (-1+6.25,0.25) rectangle (2+6.25,0.48)node[midway,text=white]{\tiny \textbf{S1}};
\draw[black,fill=orange] (2+6.25,0.25) rectangle (2.5+6.25,0.48)node[midway,text=white]{\tiny \textbf{I1}};
\draw[black,fill=teal] (2.5+6.25,0.25) rectangle (4.25+6.25,0.48)node[midway,text=white]{\tiny \textbf{S2}};
\draw[black,fill=orange] (4.25+6.25,0.25) rectangle (5.25+6.25,0.48)node[midway,text=white]{\tiny \textbf{I2}};
\draw[black,fill=teal] (11.5,0.25) rectangle (12,0.48);
\draw[black,dashed,fill=teal!50] (12,0.25) rectangle (12.5,0.48);
\draw[fill=none,draw=none] (11.5,0.25) rectangle (12.5,0.48)node[midway,text=white]{\tiny \textbf{S1}};

\draw [black,  thick ,decorate,decoration={brace,amplitude=5pt},
       xshift=5pt,yshift=-4pt] (-1.2,0.7)  -- (5.07,0.7) 
       node [black,midway,above=4pt,xshift=-2pt] {\footnotesize cycle 1};
\draw [black,  thick ,decorate,decoration={brace,amplitude=5pt},
       xshift=5pt,yshift=-4pt] (5.07,0.7)  -- (11.3,0.7) 
       node [black,midway,above=4pt,xshift=-2pt] {\footnotesize cycle 2};

\node at (0,1.2) (c0) {\texttt{conf(j1,0,0,j1\_c1)}};
\draw[thick,->] (c0)  -- (0,0.7);
\node at (5.25,1.2) (c1) {\texttt{conf(j1,1,21,j1\_c1)}};
\node at (5.23,1.5) {\texttt{step(j1,1,21)}};
\draw[->] (c1) -- (5.25,0.7);
\node at (11.5,1.2) (c2) {\texttt{conf(j1,2,46,j1\_c2)}};
\node at (11.48,1.5) {\texttt{step(j1,2,46)}};
\draw[->] (c2) -- (11.5,0.7);
\end{tikzpicture}
    \vspace*{-2mm}
\caption{Example of simulation of junction \lstinline{j1} from Figure \ref{fig:cycle}, with \lstinline{horizon=48}, \lstinline{active_p(0,stage(j1,1))}, \lstinline{active_t(0,j1,4)} and \lstinline{active_c(0,j1,j1_c1))}. S and I are shorthand representations of stages and intergreen times, respectively. } 
    \label{fig:decision_points}
\end{figure}
Figure \ref{fig:decision_points} illustrates a possible example of the predicates \lstinline{conf} and \lstinline{step}.
Our encoding can model the solution space up to a specific horizon of the three models presented in \cite{DBLP:conf/icaps/KouaitiPSMV24}, namely \textsc{CbC}, \textsc{FiRe}, and \textsc{VaRe}. Their difference regards the restrictions on forcing to keep the same configuration for at least a certain number of cycles. We define a constant \lstinline{k} representing this restriction. The rule in line $14$ detects with \lstinline{change(J,C,T,A)} the time \lstinline{T} (i.e., end of \lstinline{C}-th cycle) when the configuration for \lstinline{J} is changed to \lstinline{A}; these atoms are then used in the constraints in lines $15$ and $16$ to rule out unwanted solutions (the term \lstinline{I} in \lstinline{count_c(J,I)} is the number of  \lstinline{J}'s completed cycles at time $0$ since its configuration was changed). 

\begin{lstlisting}[float=b,label=prg:encoding2,firstnumber=17,caption={Encoding part 2 - Define \lstinline{active} predicates from time $1$ to \lstinline{horizon}}.]
time(1..horizon).
range(S,A,M,M+L-1):- available_conf(J,A), phase_limit(S,A,L), 
                     M=#sum{L1,S1: prev_status(S,S1), phase_limit(S1,A,L1)}.
active_p(T,S) :- conf(J,0,0,A), sub(J,M), range(S,A,B,E), M<E, B<=M, T=1..E-M.
active_p(T,S) :- conf(J,0,0,A), sub(J,M), range(S,A,B,E), M<B, T=B-M..E-M.
active_p(T,S) :- conf(J,C,T1,A), C>0, range(S,A,B,E), T=T1+B..T1+E, time(T).

active_t(X,J,X+M-B):- conf(J,0,0,A),sub(J,M),range(S,A,B,E),M<E,B<=M,X=1..E-M.
active_t(X,J,X+M-B):- conf(J,0,0,A),sub(J,M),range(S,A,B,E),M<B,X=B-M..E-M.
active_t(T+X,J,X-B):- conf(J,C,T,A),C>0,range(S,A,B,E),X=B..E,time(X).

active_c(1..horizon,J,A):- conf(J,0,0,A), not step(J,1,_).
active_c(1..T,J,A)      :- conf(J,0,0,A), step(J,1,T).
active_c(T..T1-1,J,A)   :- conf(J,C,T,A), step(J,C+1,T1), C>0.
active_c(T..horizon,J,A):- conf(J,C,T,A), not step(J,C+1,_), C>0, horizon>=T.
\end{lstlisting}
Once the atoms of the predicate \lstinline{conf} are defined, the rules in Listing \ref{prg:encoding2} are applied to compute the active predicates for the interval of time points ranging from $1$ to \lstinline{horizon}, entailed by \lstinline{time} in line $17$.
The rule in line $18$ defines the auxiliary atoms \lstinline{range(S,A,B,E)}, computing for each configuration \lstinline{A} the terms \lstinline{B} and \lstinline{E}, which represent the starting and ending time (relative to the cycle) for each phase \lstinline{S}. 
For instance, for the configuration \lstinline{j1_c1} in Figure \ref{fig:cycle}, we get \lstinline{range(stage(j1,1),j1_c1,0,11)}, \lstinline{range(inter(j1,1),j1_c1,12,13)}, 
\lstinline{range(stage(j1,2),j1_c1,14,20)} and 
\lstinline{range(inter(j1,2),j1_c1,21,24)}.
Then, lines $20$-$22$ define at each time \lstinline{T} the atoms \lstinline{active_p(T,P)}, containing in \lstinline{P} the active phase for each junction. 
The rules in lines $24$-$26$ define the atoms \lstinline{active_t(T,J,TS)}, containing in \lstinline{TS} the amount of time since \lstinline{P} (phase of \lstinline{J}) is active, at time \lstinline{T}, and lastly, lines $28$-$31$ define with the atoms \lstinline{active_c(T,J,A)} the active configuration \lstinline{A} for junction \lstinline{J} at time \lstinline{T}. 
Lines $20$-$21$, $24$-$25$, and $28$-$29$ compute the values for the time points included in the first cycle, while the other lines compute the values for the subsequent ones. 
The last two timelines in Figure \ref{fig:decision_points} represent the pointwise values for \lstinline{active_t} and \lstinline{active_p}; while the atoms of \lstinline{active_c} have the configuration \lstinline{j1_c1} active for every time point ranging between $1$ a $45$, and  \lstinline{j1_c2} from $46$ to $48$.

\subsection{Encoding with CASP Atoms}
\label{subsec:enc2}
\begin{lstlisting}[float=b,label=prg:encoding3,firstnumber=32,caption={Encoding part 3 - Theory atoms for occupancy and counter}.]
full(T,L,0) :- &sum{occ(T,L)}>=C, time(T), capacity(L,C).
full(T,L,1) :- time(T), link(L), not full(T,L,0).
empty(T,L,0) :- &sum{occ(T,L)}<=0, time(T), link(L).
empty(T,L,1) :- time(T), link(L), not empty(T,L,0).

in_ord(L,L1,N) :- turnrate(_,L1,L,_), N=#count{L2: turnrate(_,L2,L,_),L2<=L1}. 
out_ord(L,L1,N):- turnrate(_,L,L1,_), N=#count{L2: turnrate(_,L,L2,_),L2<=L1}.
last_in(N,L)   :- link(L), N=#count{L1: in_ord(L,L1,M)}.
last_out(N,L)  :- link(L), N=#count{L1: out_ord(L,L1,M)}.

delta(T,0,L,0) :- time(T), link(L).
delta(T,N,L,D+R*E*F) :- time(T), precedes(J,L), status(J,S), active(T-1,S), 
                        delta(T,N-1,L,D), in_ord(L,L1,N), turnrate_z(S,L1,L,R), 
                        empty(T-1,L1,E), full(T-1,L,F).
delta(T,M+N,L,D-R*E*F) :- time(T), follows(J,L), status(J,S), active(T-1,S),
                          delta(T,M+N-1,L,D), last_in(M,L), out_ord(L,L1,N), 
                          turnrate_z(S,L,L1,R), empty(T-1,L,E), full(T-1,L1,F).
&sum{D}=occ(0,L) :- initial_occ(L,D).
&sum{occ(T-1,L);D}=occ(T,L) :- delta(T,I+O,L,D), last_out(O,L), last_in(I,L).
&sum{D}=counter(0,L) :- initial_count(L,D).
&sum{counter(T-1,L);D}=counter(T,L) :- delta(T,O,L,D), last_in(O,L).
:- initial_count(L,_), &sum{counter(horizon,L) } < bound.
&maximize{counter(horizon,L) : initial_count(L,_)}.
\end{lstlisting}
To decide a promising configuration assignment, the number of PCU transiting in the corridor during the considered interval must be taken into account. To compute this information, we need to evaluate at each time point the unit of vehicles transiting in each link, considering the active stage of the adjacent junctions and the number of cars in each link (no cars can transit from one link, if its occupancy is negative, and no cars can transit in one link, if it has already reached its capacity).
Because of the size of the grounding when considering a horizon greater than $100$ seconds, it was necessary to extend the ASP encoding with CASP language (in particular, \textit{clingcon}). 
In the following, we directly describe this reformulation using  \textit{clingcon} expressions (described in Section \ref{sec:prel}).
The theory atoms consider the predicates \lstinline{occ(T,L)} and \lstinline{counter(T,L)}, associated respectively to an integer number representing the occupancy and the number of cars reaching the link \lstinline{L} at time \lstinline{T}. 
The theory atoms \lstinline|&dom{Lo..Up}=occ(T,L)| defines the minimal and maximal number that can be assigned to each \lstinline{occ(T,L)}, where \lstinline{Lo} and \lstinline{Up} are the value zero and the corresponding link's capacity (or a default value if the latter is not specified) plus an approximation error.

Listing \ref{prg:encoding3} contains the rules used to define the theory atoms and their auxiliary predicates.
The rules in lines $32$ and $33$ define the auxiliary atoms \lstinline{full(T,L,B)} where \lstinline{B} is equal to $1$ if \lstinline{L} has not reached its capacity at time \lstinline{T} and $0$ otherwise. 
Similarly, lines $34$ and $35$ define the auxiliary atoms \lstinline{empty(T,L,B)} where \lstinline{B} is $1$ if \lstinline{L} is not empty at time \lstinline{T} and $0$ otherwise. 
The rule in line $37$ defines \lstinline{in_ord(L,L1,N)} that contains, for each link \lstinline{L}, its incoming link \lstinline{L1}, where \lstinline{N} is its rank according to the lexicographic order; similarly, in line $39$, the atoms \lstinline{out_ord(L,L1,N)} contains the same information, but for the outgoing links of \lstinline{L}. 
Lines $39$ and $40$ define, respectively, \lstinline{last_in(N,L)} and \lstinline{last_out(N,L)}, two auxiliary predicates containing the number \lstinline{N} of incoming and outgoing links for each \lstinline{L}. 
All these atoms are used to compute, for each link \lstinline{L} and time \lstinline{T}, a set of \lstinline{N} atoms \lstinline{delta(T,N,L,D)}, incrementally computing the value that must be summed to the theory atoms with \lstinline{occ(T-1,L)} to obtain \lstinline{occ(T,L)}. 
Each delta starts from zero (line $42$), then, following the lexicographic order of its incoming links (thanks to the predicate \lstinline{in_ord}), it incrementally adds to the previous delta the turn rates of the current active phase, only if the incoming link is not empty and \lstinline{L} is not full (rule in line $43$). 
Note that we use the predicate \lstinline{turnrate_z}, which is not included in the listing but generalises \lstinline{turnrate} by adding the case when the turn rate is zero (i.e., not defined by the original \lstinline{turnrate}).
Lastly, line $46$ incrementally diminishes the value of delta, considering the turn rates of each outgoing link, similarly to the rule above.   
Lines $49$  define the theory atoms with \lstinline{occ} for each link at time $0$, initialising them with their initial occupancy; subsequently, line $50$ computes each link's occupancy at time \texttt{T} by adding the value of the link's last delta to its occupancy at time \texttt{T-1}.
Lines $51$ and $52$ define the theory atoms for \lstinline{counter}, which is computed similarly to \lstinline{occ}, but without considering the outgoing links (thus, it sums the value of delta obtained up to the last incoming link).
Similarly to the goal restrictions of the PDDL+ models, the constraint in line $53$ removes solutions with a value \lstinline{counter} lower than \lstinline{bound} at the time corresponding to \lstinline{horizon}. 
Additionally, we add the optimisation objective in line $54$, i.e., maximising the number of cars transiting in the links contained in  \lstinline{initial_count}. 

\section{Experiments}
\label{sec:exp}
This section describes the experiments we conducted to evaluate the performance of our approach and discusses possible applications.
In Subsection \ref{subsec:benchmark}, we detail the characteristics of the problem instances introduced in \citep{DBLP:conf/icaps/KouaitiPSMV24} considered for our experiments. 
Then, Subsection \ref{subsec:exp_details} details the experiments we run, comparing the performance of our suggested approach with the PDDL+ version. Lastly, in Subsection \ref{subsec:combination}, we highlight the optimisation capabilities of \textit{clingcon}, by suggesting possible alternative objectives and a way to combine PDDL+ with our model to guarantee that a solution is always found.

\subsection{Benchmark}
\label{subsec:benchmark}
The benchmark contains six situations in two distinct days on the corridor discussed in Section \ref{sec:desc}, and shown in Figure \ref{fig:map}: the 26th, which is a Wednesday, and the 30th, a Sunday, both in January 2022. Each day was examined at three different time slots: the morning peak hour at 8:30 am, noon at 12:30 pm, and the evening peak hour at 4:30 pm. This provides variability in terms of traffic volumes, directions, and conditions. Further, an additional situation is considered, involving exceptional traffic circumstances, i.e., a concert held at John Smith's Stadium on Tuesday the 20th of June 2023, which attracted an approximate audience of $30,000$ people. The time considered is 4:00 pm, which is before the start of the concert. This is interesting because there is a clash between commuters leaving the town and spectators arriving at the concert, creating two opposed traffic demands. For each junction, six different cycle configurations are available, generated according to historical data. 
Two different sets of cycle configurations are considered, according to the historical data used for their extrapolation, for a total of $14$ scenarios ($7\times2$).

Five instances are produced for every scenario by consistently increasing the number of links in the corridor considered in the goal, starting from \lstinline{link(wrac1,y,wrbc1)}, till considering \lstinline{link(wrec1,y,wrfc1)}. For each scenario, we identify the instances as $p_i$ for $i \in [1..5]$ where $i$ represents the number of goals.
The different numbers of goals correspond to different requirements and behaviours in the region. Focusing on one or two links in the goal can lead to flushing vehicles out of them as soon as possible, potentially congesting nearby ones. When larger chunks of the corridor are considered, there is the need to ensure traffic remains smoothly moving in the whole area. 
Overall, we consider $70$ problem instances.

\subsection{Experiment Setting and Results}
\label{subsec:exp_details}
We run our experiments on an AMD EPYC 9354 (4) $@$ 3.2GHz machine with $32$ GB of memory under Linux (Ubuntu 22.04.5 LTS), using the system \textit{clingcon} (v5.2.1) with the \textit{libclingo} v5.8.0. 
To run the PDDL+ solver, \textit{enhsp} \citep{enhsp}, we use \textit{java openjdk} (v21.0.6).
The instances, ASP and \textit{clingcon} encoding, as well as the experiment setup can be found at this link \href{https://github.com/altarzariol/traf_sign_casp}{https://github.com/altarzariol/traf\_sign\_casp}.

Our \textit{clingcon} encoding models the same solution space (up to a certain horizon) of the  PDDL+ models from \cite{DBLP:conf/icaps/KouaitiPSMV24}, in particular, in the experiments we focus on \textsc{FiRe} (i.e, setting the limit $k=4$ for every junction) since it is the model obtaining the best performance among the proposed PDDL+ alternatives and that, on average, showed better plans than the ones used in historical data. 
In the paper, the authors set a threshold of $350$ vehicles in every goal, and evaluate the plans found by cutting their horizons up to $900$ seconds (i.e., $15$ minutes). Limiting the plan's horizon for the evaluation was necessary in order to consider only simulations consistent with the real-world; indeed, when considering horizons greater than $15$ minutes, the results diverge from the simulations because of the shifting of underlying turn rates \citep{DBLP:conf/icaart/BhatnagarMSMMV22}.
Therefore, for our \textit{clingcon} encoding, we set the constant \lstinline{horizon} up to $900$. 
Moreover, while state-of-the-art PDDL+ approaches for traffic signal optimisation can generate solutions quickly, in many cases, strategies are generated in advance, and validated and tested before being deployed in the controlled region -- effectively enabling the use of approaches that could generate higher quality solutions more slowly. 
For this reason, in the following experiments, we set a timeout of $10$ minutes for every run.

To validate the simulation obtained with our \textit{clingcon} encoding, we compare the status of the corridor and decision points with respect to \textsc{FiRe}, using the \texttt{pps} simulation tool \href{https://github.com/hstairs/pps}{https://github.com/hstairs/pps}: this approach demonstrated to support simulations that are close to real-world traffic evolution within $15$ minutes time windows \citep{bhatnagar2022leveraging}.
We evaluate two tasks: in the former, we compare our encoding to the PDDL+ model \textsc{FiRe} on a similar setting, i.e., the decision version of the problem; while, in the latter, we compare the result of \textsc{FiRe} with the optimisation version of our \textit{clingcon} encoding. For both tasks, we run \textit{clingcon} with the flag \lstinline{--config=crafty} since this is the option that obtained the best performance in our experiments.
\begin{figure}
    \centering
    \includegraphics[width=0.5\linewidth]{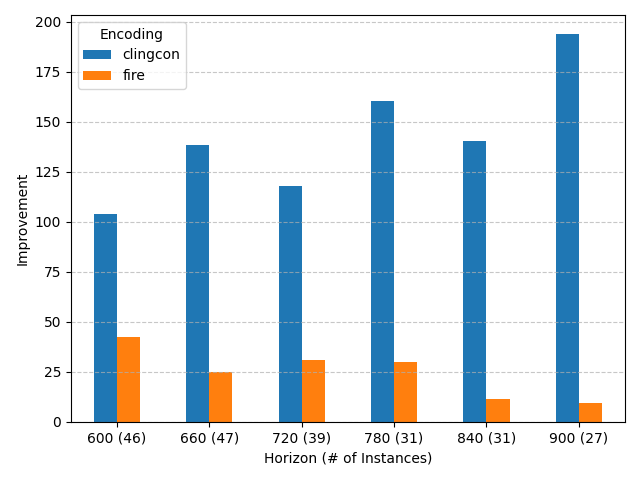}
    \caption{Task 1 - Decision version with bound.}
    \label{fig:task1}
\end{figure}
\paragraph{Task 1: Decision Problem with Bound.}
The horizon of the plans returned by \textsc{FiRe} with goals of $350$ vehicles is higher than $15$ minutes. Thus, to provide a fair ground to compare the two approaches, we set the constant \lstinline{bound} for each instance in the benchmark by considering the minimal counter observed in the plans of \textsc{FiRe} at $10,.., 15$ minutes (namely, $600,.., 900$ seconds) and setting the constants \lstinline{bound} and \lstinline{horizon} accordingly.
Figure \ref{fig:task1} shows the aggregated improvement of the values  \lstinline{counter} in the corresponding horizon, considering only the instances for which \textit{clingcon} managed to find a plan within a timeout of $10$ minutes. In the axis ``Horizon", we report the value \lstinline{horizon} used by \textit{clingcon} and the limit used for the evaluation of the plan found by \textsc{FiRe}, while in parentheses we specify the number of considered instances (i.e, those for which \textit{clingcon} found a solution equal or better than the given bound within the timeout). 
The results indicate that, despite the struggle with the grounding size,  \textit{clingcon} can be used to improve PDDL+ solutions. Moreover, although for $43$ instances with an horizon of $900$ seconds we get a timeout, the solutions found for the remaining $27$ instances obtained a considerable improvement. 

\paragraph{Task 2: Optimisation Problem without Bound.}
Subsequently, we evaluate our encoding by running it with the optimisation statement in line $54$ of Listing \ref{prg:encoding3}. 
We set a time limit of $10$ minutes, no request on the bound (thus, \lstinline{bound=0} and \lstinline{horizon=900}), and the results considered are derived from the best solution obtained within the time limit. 

\begin{figure}%
    \centering
    \subfloat[\centering Aggregated results projecting horizon.]{{\includegraphics[width=6.5cm]{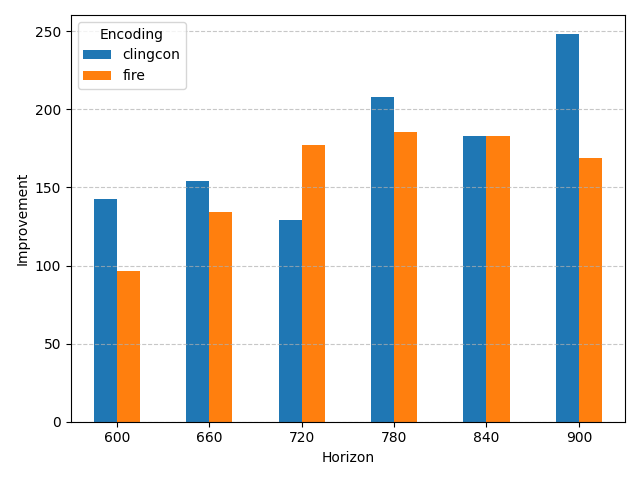} \label{fig:task2a}}}%
    \quad
    \subfloat[\centering Aggregated results at horizon 900.]{{\includegraphics[width=6.5cm]{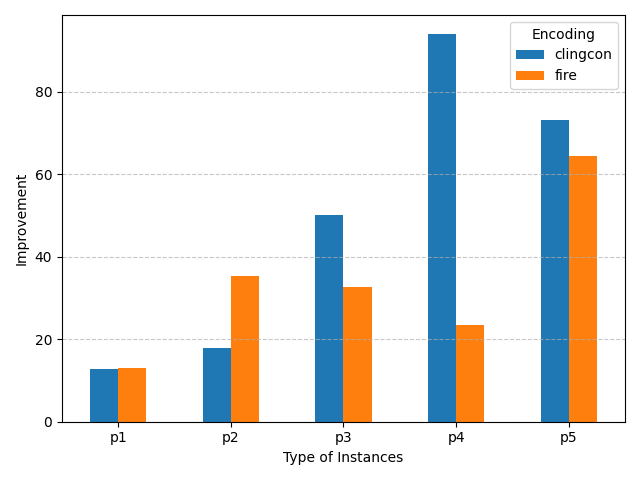} \label{fig:task2b}}}%
    \caption{Task 2: Optimisation version without bound.}%
    \label{fig:task2}%
\end{figure}

Figure \ref{fig:task2a} shows the aggregated improvements observed for the $70$ instances of the benchmark, by projecting the plan of the two approaches at different time points. Although not every \textit{clingcon}'s plan improves the quality of the solutions returned by \textsc{FiRe}, the overall improvement from its successful runs overcomes the one of \textsc{FiRe}.
The only exception can be seen at $720$ seconds. One possible explanation for this observation is that, in that moment or slightly before, most of  \textsc{FiRe} solutions maximise the flow in the corridor, possibly penalising neighbouring links, and subsequently leaving the corridor with low occupancy, thus requiring more time to increase the counter consistently. 
Figure \ref{fig:task2b}, on the other hand, projects the aggregated result at horizon $900$, dividing it by type of instances.
Here we can observe that the majority of \textit{clingcon} improvements derive from instances with three, four or five goals. One reason for this observation is that, by focusing on one or two links, it is easier to congest the link below, and then get worse overall results.
Lastly, in both tasks, we observe that \textit{clingcon} struggles with returning a solution when many decision points occur in the simulated horizon. 

\subsection{Alternative Optimisations and Combination}
\label{subsec:combination}
Thanks to its optimisation capabilities, \textit{clingcon}  can express alternative optimisation goals, making the suggested model modular and capable of adapting to different targets. 

Indeed, one may want to increase the occupancy of a link as much as possible -- a proxy for reducing vehicles' speed on that link. 
Another goal can be to reduce the occupancy of a link as much as possible, to optimise the flow of traffic from a specific entry point/direction. 
In both cases, the most effective way to indirectly express the optimisation statement is to maximise/minimise the values computed with the predicates \lstinline{delta}, which represent the increment/decrement in the occupancy of a link at every time. This allows for reducing the size of the values considered, which can quickly become intractable when considering the difference between occupancy and capacity. 
To do that, we can simply add the following \textit{clingcon} expressions to define a new theory atom \lstinline{increments(T,L)} that represents the total PCU that entered in/exited from the link \lstinline{L} identified by the input atoms \lstinline{max_occupancy(L)} (if we consider maximisation) at time \lstinline{T}:

\lstinline|&sum{0}=increments(0,L):-  max_occupancy(L).|

\lstinline|&sum{increments(T-1,L);D}=increments(T,L):-| 

\lstinline|   delta(T,I+O,L,D), last_delta_out(O,L), last_delta_in(I,L), max_occupancy(L).|

\noindent
Then, we maximise the occupancy of the target links with the following optimisation statement:

\lstinline|&maximize{increments(horizon,L) : max_occupancy(L)}.| 

\noindent
If, on the other hand, we aim to minimise the occupancy of target links, we can simply add the same rules but instead of \lstinline{max_occupancy(L)}, we represent the target with \lstinline{min_occupancy(L)} and then, instead of \lstinline{&maximize} we write an optimisation statement with \lstinline{&minimize}.

It is also possible to encode goals where counters on exit links have to be minimised -- to represent cases where we want to reduce traffic pressure on a nearby region by slowing down vehicles in the controlled one. 
Similarly to the previous case, to achieve this, we can simply replace the directive \lstinline{&maximize} with \lstinline{&minimize} in line $54$ of Listing \ref{prg:encoding3}.  

Lastly, more complicated goals can be used to represent traffic accident management, where traffic before the accident has to be slowed down (minimise counter value), while traffic on that link and subsequent ones has to be flushed away as soon as possible (maximise corresponding counters).
To model this, we can identify the links before and after the accident, e.g., with \lstinline{slow_traffic(L)} and \lstinline{flush_traffic(L)}, and write the following optimisation statements:

\lstinline|&maximize{counter(horizon,L) : flush_traffic(L)}.|

\lstinline|&minimize{counter(horizon,L) : slow_traffic(L)}.|

\noindent
As long as all these optimisation statements do not aim to pursue contrasting objectives, e.g., by specifying the same link for \lstinline{max_occupancy(L)} and \lstinline{min_occupancy(L)}, we can include more than one optimisation target in the encoding and even specify the priority level, similarly to weak constraints for ASP.

\paragraph{Combining PDDL+ and CASP.}
The \textit{clingcon} encoding can also be used to improve the quality of solutions returned by the PDDL+ approach, hence combining the strengths of the approaches. From the solution found by the PDDL+ planner, the values of \lstinline{counter} at a given horizon of each target link can be extracted. This information can be encoded in atoms of the form \lstinline{pddl_solution(L,C)}, where \lstinline{L} is a target link for the optimisation and \lstinline{C} is the value of \lstinline{counter} obtained by the PDDL+ plan. Then, by including in our encoding the following constraint, we can force \textit{clingcon} to return a solution that is strictly better than the PDDL+ one:

\lstinline|:- &sum{counter(horizon,L) : pddl_solution(L,_)} <= S,|

\lstinline|   S=#sum{B,L: pddl_solution(L,B)}, pddl_solution(_,_).|

\noindent
We implemented an automated pipeline to exploit the synergies of the two approaches and tested it on the same benchmark and settings used for our experiments. We considered horizons of $10,..., 15$ minutes. 
By combining PDDL+ with ASP, we managed to improve the quality of almost half of the solutions. 
The automated pipeline and the results of the experiments can be found at this link \href{https://github.com/altarzariol/traf_sign_casp}{https://github.com/altarzariol/traf\_sign\_casp}.

\section{Related Work}
\label{sec:rel}

A large number of planning and scheduling-based approaches have been developed for traffic signal optimisation. \citeauthor{promet} (\citeyear{promet}) proposed a system that integrates an AI planning engine with the \textsc{Sumo} simulator \citep{SUMO2018} via an ``Intelligent Autonomic System'' module. Their \textsc{pddl2.1} model utilises relative density descriptors (e.g., ``low'', ``medium'') to represent traffic concentration on road links, abstracting away from individual vehicle counts. This approach enables scalability to regions with thousands of vehicles. 
The work by \citeauthor{pozanco2021line} (\citeyear{pozanco2021line}) builds upon this approach, introducing also the ability for continuous learning and knowledge model evolution for improved network adaptation. 
The preliminary work by \citeauthor{DBLP:conf/socs/IvankovicVCR22} \citeyear{DBLP:conf/socs/IvankovicVCR22} performs traffic signal optimisation by leveraging on planning techniques that reason with global state constraints \citep{DBLP:journals/jair/HaslumIR0TSN18}, which can provide valuable insights into the broader impact of light changes.

On a different line of work \citep{vallati2016efficient,mccluskey2017embedding} exploits PDDL+ for encoding a flow model of vehicles through traffic-light controlled junctions. Those initial works have then been extended in \citep{percassi2023practical,DBLP:conf/icaps/KouaitiPSMV24}, where the proposed approaches have been extensively validated with historical data from urban regions in Manchester and Huddersfield, from the northern part of the United Kingdom. These most recent works take into account the constraints of the existing infrastructure and are hence suitable for deployment. Finally, \cite{percassi2025automated} assesses the suitability of LLMs to generate valid and effective traffic signal configurations from scratch, and  \cite{percassi2024leveraging} demonstrates how to perform a what-if analysis on the basis of the engineered knowledge models. 

From a different perspective, the SurTrac system leverages a decentralised scheduling technique for urban traffic signal control \citep{Smith12,DBLP:conf/aips/HuS19,DBLP:journals/aim/Smith20}. Each intersection acts as an autonomous scheduling agent, collaborating with neighbouring intersections to predict future traffic demand and minimise expected vehicle wait times at their respective signals. This distributed approach exhibits good potential for scalability due to its localised decision-making, but may exhibit reduced flexibility in achieving specific system-wide goals compared to centralised methods. 

Instead, there are far fewer approaches to problems related to traffic signal optimisation that use ASP. 
\cite{DBLP:conf/ecai/EiterFSS20} introduces an approach to optimise the coupling of traffic movements at junctions, according to expected traffic demand in the area, and to simulate using a mesoscopic-level representation. 
The experiments considered a realistic area with two junctions and are compared to the microscopic traffic simulator SUMO \citep{SUMO2018}. In the wider area of traffic control, \cite{DBLP:journals/tplp/CardelliniDMV24} deals with the problem of dynamic traffic distributions in urban areas. As a part of a framework defined for solving such problem, they employed ASP for the computation of the best possible routes for all the vehicles in the network, starting from a set of candidate routes for all vehicles within the framework. On other directions in traffic research, ASP has been employed by \cite{DBLP:conf/aaai/BeckEK12,DBLP:conf/jelia/BeckEK12} to manage the inconsistency in traffic regulations in Smart Cities, and by \cite{DBLP:conf/ai/VaseqiD13} 
as a component in a situation awareness system for maritime traffic control.

\section{Conclusion}
\label{sec:conc}
In this paper, we presented a novel approach to the traffic signal optimisation problem leveraging CASP. By encoding the problem within a bounded time horizon, our method addresses a key limitation of existing PDDL+ solutions, which do not properly support the specification of optimisation criteria. Our empirical evaluation, conducted on real-world historical traffic data for a range of traffic conditions, highlighted the capabilities of the proposed approach and its benefits over the PDDL+ state of the art, as well as the potential of combining the approaches. 
Future work will focus on further improving the solving phase by tackling the limitations of one-shot search by adapting multi-shot techniques, and by defining and implementing domain heuristics, possibly extending the approach to larger urban regions. We are also interested in exploring the use of ASP-based approaches for identifying suitable traffic signal cycle configurations to be used according to the expected traffic conditions to be dealt with. 


\subsubsection*{Acknowledgments} 
We thank the anonymous reviewers for their valuable feedback and suggestions that helped us extend the paper's contributions. We also thank Martin Gebser for his suggestions on the ASP encoding. 
Mauro Vallati was supported by a UKRI Future Leaders Fellowship [grant number MR/Z00005X/1].
Marco Maratea was supported by the European Union - NextGenerationEU and by Italian Ministry of Research (MUR) under PNRR project FAIR ``Future AI Research'', CUP H23C22000860006. This research was funded in part by the Austrian Science Fund (FWF) 10.55776/COE12. For open access purposes, the author has applied a CC BY public copyright license to any author accepted manuscript version arising from this submission.

\bibliographystyle{tlplike}

\end{document}